\documentclass[9pt,twocolumn,twoside]{osajnl}

\journal{ol_noheader} 

\setboolean{shortarticle}{true}

\newcommand{\work}{Letter}

\newcommand{\Er}{\relax\ifmmode E_\mathrm{ref} \else $E_\mathrm{ref}$\ \fi}
\newcommand{\Ep}{\relax\ifmmode E_\mathrm{probe} \else $E_\mathrm{probe}$\ \fi}
\newcommand{\Ee}{\relax\ifmmode E_\mathrm{empty} \else $E_\mathrm{empty}$\ \fi}
\newcommand{\tEr}{\relax\ifmmode \tilde{E} \else $\tilde{E}$\ \fi}

\title{Imaging trapped quantum gases by off-axis holography}

\author[1,*]{J. Smits}
\author[1]{A.\,P. Mosk}
\author[1]{P. van der Straten}

\affil[1]{Debye Instistute for Nanomaterials Physics and Center for Extreme Matter and Emergent Phenomena, Utrecht University, Princetonplein 1, 3584 CC, Utrecht, The Netherlands}

\affil[*]{Corresponding author: j.smits@uu.nl}

\ociscodes{(020.1475) Bose-Einstein condensates; (070.0070) Fourier optics and signal processing; (090.1995) Digital holography; (120.3180) Interferometry; (120.5050) Phase measurement}

\doi{}

\begin{abstract}
We present a dispersive imaging method for trapped quantum gases based on digital off-axis holography. Both phase delay and intensity of the probe field are determined from the same image. Due to the heterodyne gain inherent to the holographic method it is possible to retrieve the phase delay induced by the atoms at probe beam doses two orders of magnitude lower than phase-contrast imaging methods. Using the full field of the probe beam we numerically correct for image defocusing.
\end{abstract}

\setboolean{displaycopyright}{false}

\begin{document}

\maketitle

When studying the dynamics of trapped quantum gases, it is desirable to have a method of imaging that perturbs the atom cloud as little as possible, which makes it possible to perform a study on atom cloud dynamics on a single sample. Due to the extremely low temperatures of quantum gases any photon absorption event induces significant atom losses, which influences the outcome of a sequence of measurements. To reduce scattering the frequency of the light can be detuned from the atomic transition, but this in turn reduces the refractive index contrast. This makes that a quantum gas has at the same time very low refractive index contrast and can endure very little probe light. In this \work\ we present a dispersive imaging method for quantum degenerate atom clouds based on off-axis holography \cite{Leith62,Takeda82,Cuche00,Park18}. In addition to the probe beam, a reference beam is used which interferes with the probe beam. From the interference pattern between the two beams, the full field of the probe beam is reconstructed. Use of an external reference beam enables imaging at the probe beam shot noise level for any intensity. As atom losses are directly related to the dose of a light pulse, reducing the intensity or pulse time of the probe beam reduces the atom losses allowing for longer interrogation time of the same sample. Moreover, as the recorded hologram contains both absorption and phase delay resulting from interaction with the atoms, both can be studied independently. Since the full field of the probe beam is known, it is possible to use numerical refocusing \cite{Ferraro05} to correct for defocusing in the experiment using data post-processing.

Many different imaging methods have been developed \cite{Ramanathan12,Andrews96,Gajdacz13,Wigley16,Light13,Turner05,Sobol14} to image quantum gases at minimum losses, most notably partial-transfer absorption imaging \cite{Ramanathan12} and several dispersive methods such as phase-contrast imaging \cite{Andrews96}, and more recently dark-field Faraday rotation imaging \cite{Gajdacz13} and shadowgraph imaging \cite{Wigley16}. However, these dispersive methods lack the heterodyne gain present in a method based on off-axis holography leading to increased noise at lower probe doses. Holographic imaging methods have been demonstrated on atoms trapped in a magneto-optical trap \cite{Kadlecek01,Sobol14} and in an optical lattice \cite{Hoffmann16}.

In quantum gases the refractive index is proportional to the density \cite{Meppelink10}. By changing the detuning $\delta$ from the atomic resonance a quantum gas can be made mostly an absorber ($\delta \approx 0$, on resonance) or mostly a phase-object ($|\delta|\gg\gamma$, many atomic linewidths $\gamma$ detuned). The imaginary part of the refractive index, responsible for absorption, scales as $\Im(\mathcal{N}) \propto 1/\delta^2$ for large detuning $\delta$. The real part, responsible for the phase shift, scales as $\Re(\mathcal{N}) \propto 1/\delta$. Increasing the detuning from resonance reduces atom losses, at the cost of signal strength. Experiments are performed at $\delta = -350\,\mathrm{MHz}$ or approximately 36 atomic linewidths from the resonance. At typical peak densities in these experiments, the refractive index contrast is $\Re(\mathcal{N})-1 \approx 2\times10^{-3}$.

\begin{figure}[htb]
\centering
\includegraphics{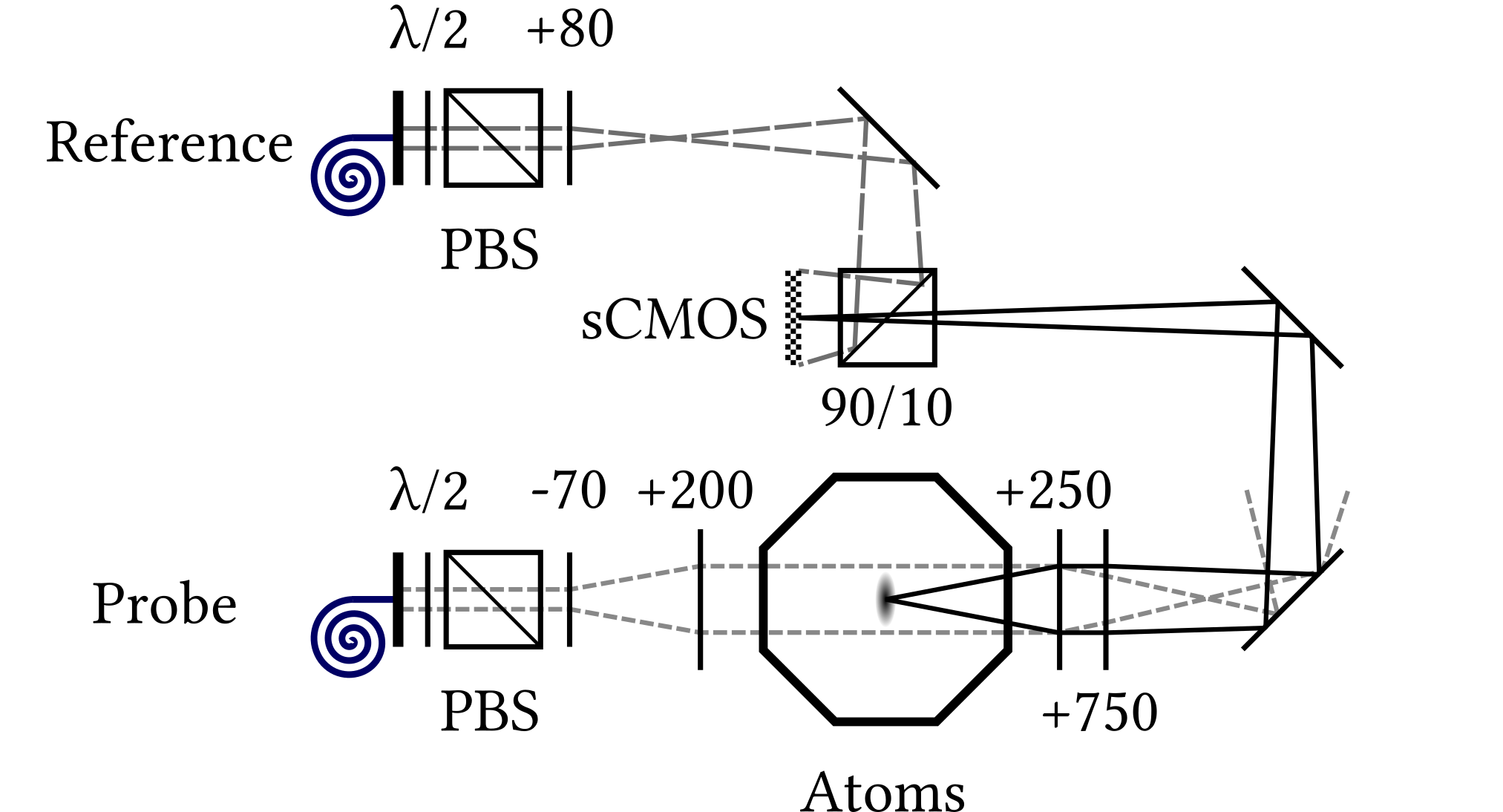}
\caption{Schematic representation of the setup. The focal distance of lenses is given in millimetres. The octagon represents the vacuum chamber and is approximately 50\,cm across. After each fiber there is a half-wave plate, denoted by $\lambda/2$, and a polarizing beam splitter, denoted by PBS, to ensure the polarization of the light. The label sCMOS denotes the camera.}
\label{fig:setup}
\end{figure}

The imaging is performed on a Bose-condensed gas of Na atoms. The atoms are trapped in a cylindrically-symmetric magnetic trap with effective trapping frequencies $(\omega_\rho,\omega_z) = 2\pi\times(60.0,15.0)\,\mathrm{Hz}$ and are cooled to below the critical temperature for Bose-Einstein condensation by means of evaporative cooling, reaching temperatures between $400$ and $600\,\mathrm{nK}$ with approximately $N = 5\times10^{7}$ particles. The atoms are illuminated by a locally flat beam and are imaged on the camera (see Fig.~\ref{fig:setup}). To be able to perform off-axis holography, the probe light is split on a separate optical table and both probe and reference beam are transported to the experimental setup by polarization maintaining fibers. The reference beam is matched in divergence to the unscattered probe light to cancel the relative curvature between the wavefronts of probe and reference beam. The angle between the probe and reference beam on the camera is a few degrees. The large size of the vacuum chamber limits the numerical aperture using optics with a $50$-$\mathrm{mm}$ diameter to $\mathrm{NA} = 0.1$. Due to the use of two separate fibers to transport reference and probe beam a global phase shift will be present for each image, which is removed in post-processing by setting the part of the image where no atoms are present to zero accumulated phase. The coherence length of the probe laser is in the order of $100\,\mathrm{m}$ owing to the $1\,\mathrm{MHz}$ laser linewidth. Since probe illumination time of individual images are in the order of $10-100\,\mu\mathrm{s}$, no special precautions are taken to ensure stability from vibrations, as the probe pulse is generally shorter than the time scales of vibrations in the setup. Since a typical cold atoms laboratory has lasers operating at sub-$\mathrm{MHz}$ laser linewidth, adding a reference beam on the camera at a small angle with respect to the probe beam is all that is needed to convert an existing imaging system for absorption or phase-contrast imaging to use this holographic method.

\begin{figure}[htb]
\centering
\includegraphics{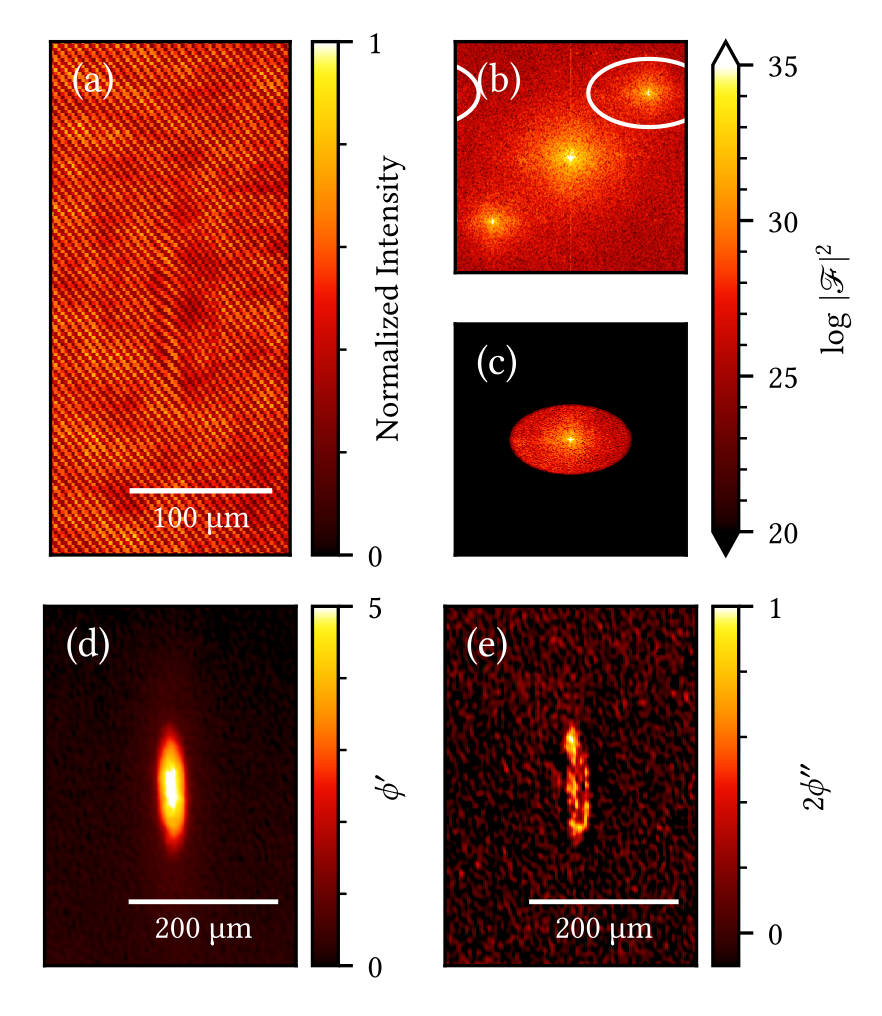}
\caption{From hologram to phase and optical density. \emph{(a)} Cut-out of the interference pattern as recorded by the camera, centered on the atom cloud. Note that the fringes are curved due to the extra accumulated phase. \emph{(b)} Fourier space before the cut and translation. The cut-out is indicated by the white ellipse. \emph{(c)} Fourier space after cut and translation. \emph{(d)} Accumulated phase extracted from the inverse Fourier transform, centered on the atom cloud. \emph{(e)} Optical density extracted from the inverse Fourier transform centered on the atom cloud.}
\label{fig:holography}
\end{figure}%

The resulting image on the camera is an interference pattern between probe and reference beam,%
\begin{multline}
    I \propto  |\Er e^{i \mathbf{k}_\mathrm{ref}\cdot\mathbf{r}} + \Ep e^{i \mathbf{k}_\mathrm{probe}\cdot\mathbf{r}} |^2 \\=|\Er|^2 + |\Ep|^2 + \Er^* \Ep e^{i \tilde{\mathbf{k}}\cdot\mathbf{r}} + \Er \Ep^* e^{-i \mathbf{\tilde{k}}\cdot\mathbf{r}},
    \label{eq:image}
\end{multline}%
where $\mathbf{r} = (x,z)$ and $\mathbf{\tilde{k}} = \mathbf{k}_\mathrm{probe}-\mathbf{k}_\mathrm{ref}$ is the difference wavevector of the incoming fields, which is determined by the angle $\theta_x,\theta_z$ between the reference and probe beam, and given by $\mathbf{\tilde{k}} = k_0 ( \sin \theta_x , \sin \theta_z )$, where $k_0$ is the laser wavenumber. A cutout of such an intensity profile, centered on the atom cloud, is shown in Fig. \ref{fig:holography}a. The Fourier transform of the intensity pattern contains well-defined peaks associated with the interference pattern. To prevent artefacts in the fast Fourier transform due to boundary effects, a square Tukey window with width $\alpha=0.1$ is applied prior to applying the Fourier transform. The result of the Fourier transform  is shown in Fig. \ref{fig:holography}b. Focussing on one of the interference terms in Eq. \ref{eq:image}, the Fourier transform is given by%
\begin{equation}
    \mathcal{F}(\Er^* \Ep e^{i \tilde{\mathbf{k}}\cdot\mathbf{r}})(\mathbf{k}) = \mathcal{F}(\Er^* \Ep)(\mathbf{k}-\tilde{\mathbf{k}}).
    \label{eq:fourier}
\end{equation}%
Essentially, the interference term in the Fourier transform contains the information of the product of the fields translated from the origin. By taking an appropriate cutout in Fourier space, the information of the product of these fields can be isolated. In this case, an elliptical window (Tukey, $\alpha=0.1$) is chosen, resulting in numerical apertures $\mathrm{NA}_x = 0.064$ and $\mathrm{NA}_z = 0.040$ (see Fig. \ref{fig:holography}b). In Fig. \ref{fig:holography}c the cutout has been translated to the origin. 

An elliptical window is chosen as the elongated shape of the atom cloud will have a larger extent in Fourier space in its short ($x$) direction, while being relatively compact in Fourier space in the long ($z$) direction. This yields the most accurate image of the atom cloud at rest, but the choice of window shape should be considered based on the type of experiment that is performed. The inverse Fourier transform of the cutout yields the product of the fields of the probe and reference beam. The reference beam is sufficiently flat, such that the result is the field of the probe beam, scaled by the magnitude of the field of the reference beam. By applying an inverse Fourier transform to the cutout the full field, both amplitude and phase, of the probe beam can be retrieved. For normalization, a second recording without atoms is made afterwards to calculate the normalized field of the probe beam,%
\begin{equation} 
    \tEr =  \frac{\Er^* \Ep}{\Er^*\Ee} \equiv e^{-\phi'' - i \phi'}, 
    \label{eq:tEr} 
\end{equation}%
where $\phi'$ is the phase delay of the probe beam accumulated as it passes through the atom cloud, and $2\phi''$ is the optical density. The argument of $\tEr$ is directly proportional to the phase delay, as opposed to phase contrast imaging or shadowgraph imaging methods in which the phase is reconstructed from the intensity profile, which makes these methods more susceptible to noise. The phase delay is shown in Fig. \ref{fig:holography}d. The Bose-Einstein condensate is seen as a dense core in a diffuse thermal cloud. The optical density is extracted from the field amplitude and is shown in Fig. \ref{fig:holography}e. Here the signal from the thermal cloud is too weak to be observed, but the Bose-Einstein condensate is clearly visible. This signal is very dependent on the chosen focal plane. The absence of signal for the thermal cloud in this image is due to the much lower density in the thermal cloud. To determine the density distribution in the atom cloud for $|\delta|\gg\gamma$, it is sufficient to consider only the accumulated phase.

\begin{figure}[htb]%
\centering%
\includegraphics{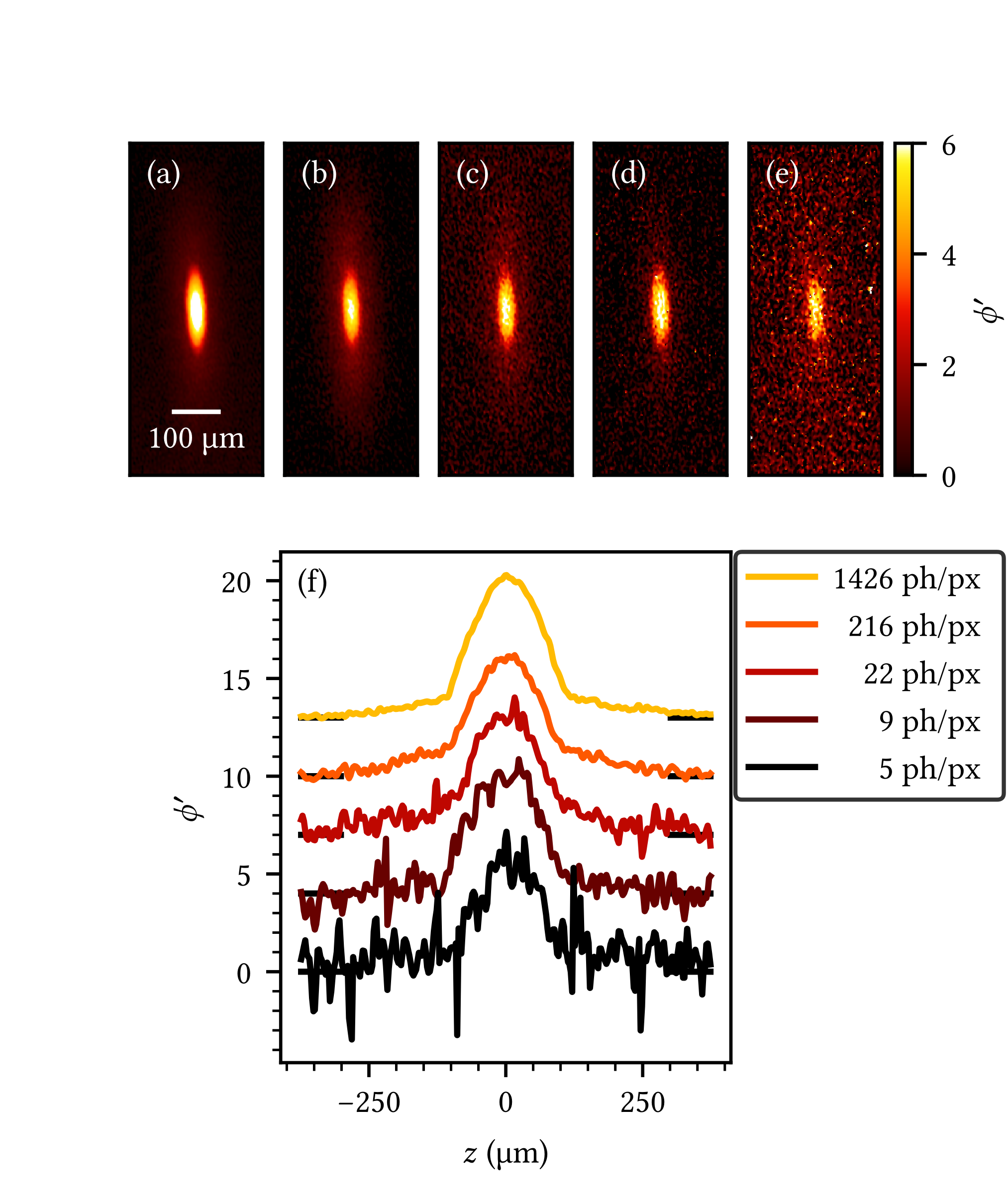}
\caption{Advantages of heterodyne gain in determining accumulated phase. All doses in irradiated photons per pixel. \emph{(a-e)} Degenerate atom clouds imaged at different photon doses at similar atom numbers in the atomic cloud. \emph{(a)} Pulse time of $\tau = 100\,\mu\mathrm{s}$ at intensity $I = 69\,\mu\mathrm{W}/\mathrm{cm}^2$, corresponding to $1426\,\mathrm{photons}/\mathrm{px}$. \emph{(b)} $\tau = 50\,\mu\mathrm{s}$, $I = 21\,\mu\mathrm{W}/\mathrm{cm}^2$, $216\,\mathrm{photons}/\mathrm{px}$. \emph{(c)} $\tau = 25\,\mu\mathrm{s}$, $I = 4\,\mu\mathrm{W}/\mathrm{cm}^2$,  $22\,\mathrm{photons}/\mathrm{px}$. \emph{(d)} $\tau = 10\,\mu\mathrm{s}$, $I = 4\,\mu\mathrm{W}/\mathrm{cm}^2$, $9\,\mathrm{photons}/\mathrm{px}$. \emph{(e)} $\tau = 5\,\mu\mathrm{s}$, $I = 4\,\mu\mathrm{W}/\mathrm{cm}^2$, corresponding to $5\,\mathrm{photons}/\mathrm{px}$. \emph{(f)} Slice through a single row of pixels in the center of the cloud for figures \emph{(a-e)}. The black horizontal lines indicate the zero level for the different lines.}
\label{fig:fig_lowlight}
\end{figure}

For minimally destructive imaging the dose of the probe beam is chosen as low as possible, while preserving a sufficiently low noise level for analysis of the resulting image. Since the reference beam can be chosen arbitrarily intense for increased heterodyne gain, off-axis holography allows shot-noise limited imaging down to the single-count-per-pixel level in the reference beam \cite{Verpillat10}. In Fig.~\ref{fig:fig_lowlight} we demonstrate the effect of reduced probe power and duration on the image quality. Figure~\ref{fig:fig_lowlight}a approximately corresponds to typical probe power and duration used in phase-constrast imaging in earlier experiments \cite{Bons16}, which is an irradiated dose of approximately 1700 $\mathrm{photons}/\mathrm{px}$. At lower probe power we observe an acceptable increase in noise down to an irradiated dose of 9 $\mathrm{photons}/\mathrm{px}$. At an irradiated dose of 4 $\mathrm{photons}/\mathrm{px}$ the reconstructed field contains phase vortices, as can be seen in the cut-through in Figs.~\ref{fig:fig_lowlight}e and \ref{fig:fig_lowlight}f. These spurious vortices can be attributed to a reduction in fringe contrast and decrease the signal-to-noise ratio in the reconstructed phase. A slice through each cloud along its long axis is shown in Fig.~\ref{fig:fig_lowlight}f.

\begin{figure}[htb]
\centering
\includegraphics{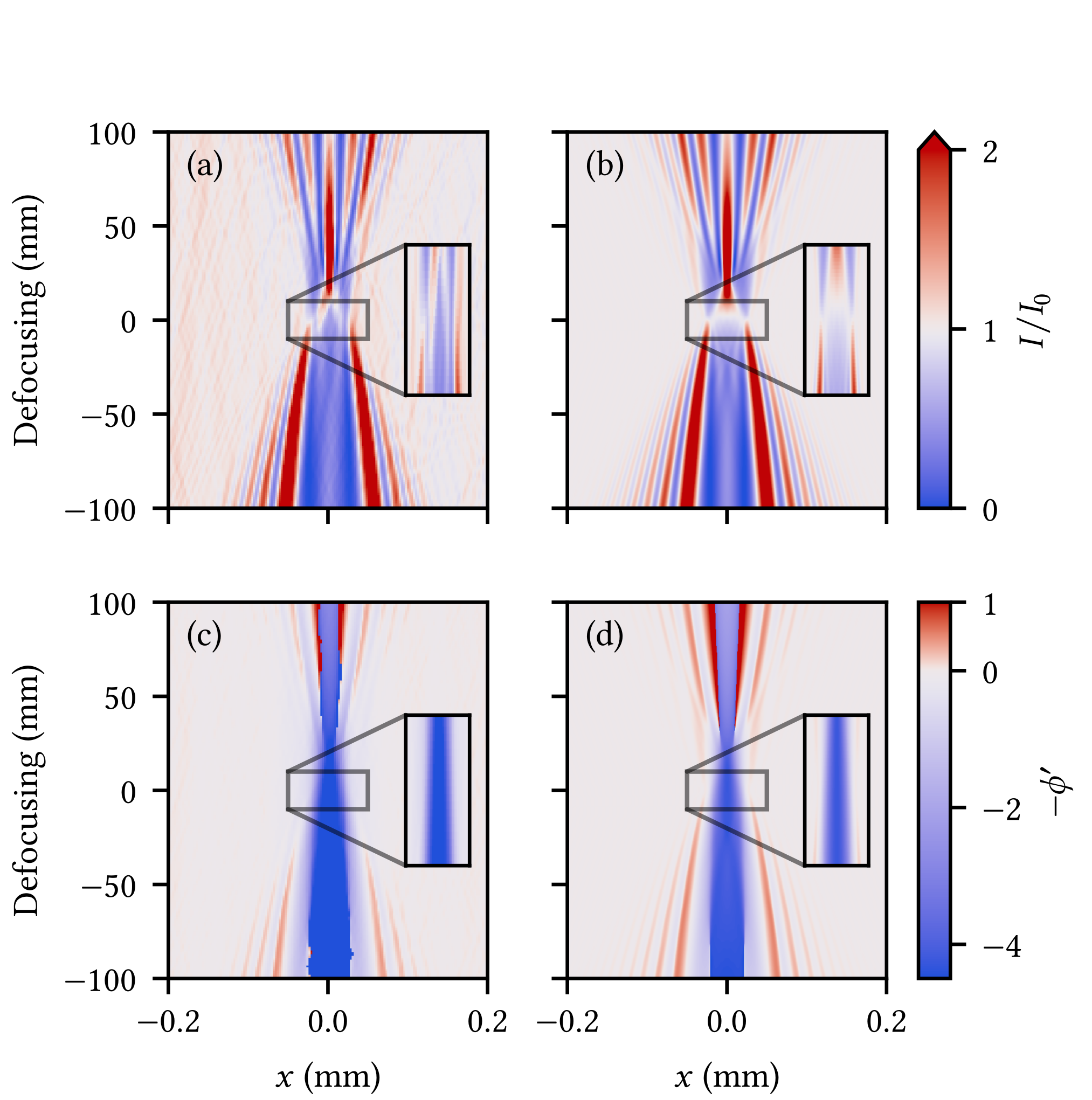}
\caption{Effect of defocusing on phase and amplitude. Cross-section through the center of the atomic cloud in the radial direction. The total span of the inset is $1\,\mathrm{cm}$. \emph{(a)} Effect of numerical defocusing on the intensity in a typical experiment. \emph{(b)} Comparison of (a) to a numerical model. \emph{(c)} Effect of numerical defocusing on the phase in a typical experiment. \emph{(d)} Comparison of (c) to a numerical model.}
\label{fig:focus}
\end{figure}

To study the effect of defocusing on the quality of the image, a slice of the field is taken through the radial direction of the condensate. Since the full field is known, the Beam Propagation Method (BPM) \cite{Thylen83,goodman} in free space can be used. With the BPM the field is calculated at different planes, such that to propagate the field from a plane at $y$ to a plane at $y'$ one calculates %
\begin{equation}
    \tEr(x,y') = \mathcal{F}^{-1}_x \left\{ e^{ -i k^2 (y' - y)/(2k_0) } \times \mathcal{F}_x[\tEr(x,y)](k) \right\}(x)  .
    \label{eq:refocus}
\end{equation}%
For comparison, light propagation through a Bose-Einstein condensate is calculated using the BPM and a time-splitting spectral method \cite{Bao02}. In the calculation, a cut is made in Fourier space to simulate $\mathrm{NA}_x = 0.064$, as in the experiment. The results of propagating the experimental results and the comparison to theory are shown in Fig. \ref{fig:focus}. The intensity varies strongly at slight defocusing due to cross-talk between phase and amplitude. This can be attributed to the lensing properties of the Bose-Einstein condensate: Light passing through the atom cloud is refracted causing strong dependence on the chosen focus plane. The phase is more robust against defocusing yielding an interpretable signal even at slight misalignment of the focus, but defocusing will change the perceived dimensions of the cloud.

\begin{figure}[htb]
\centering
\includegraphics{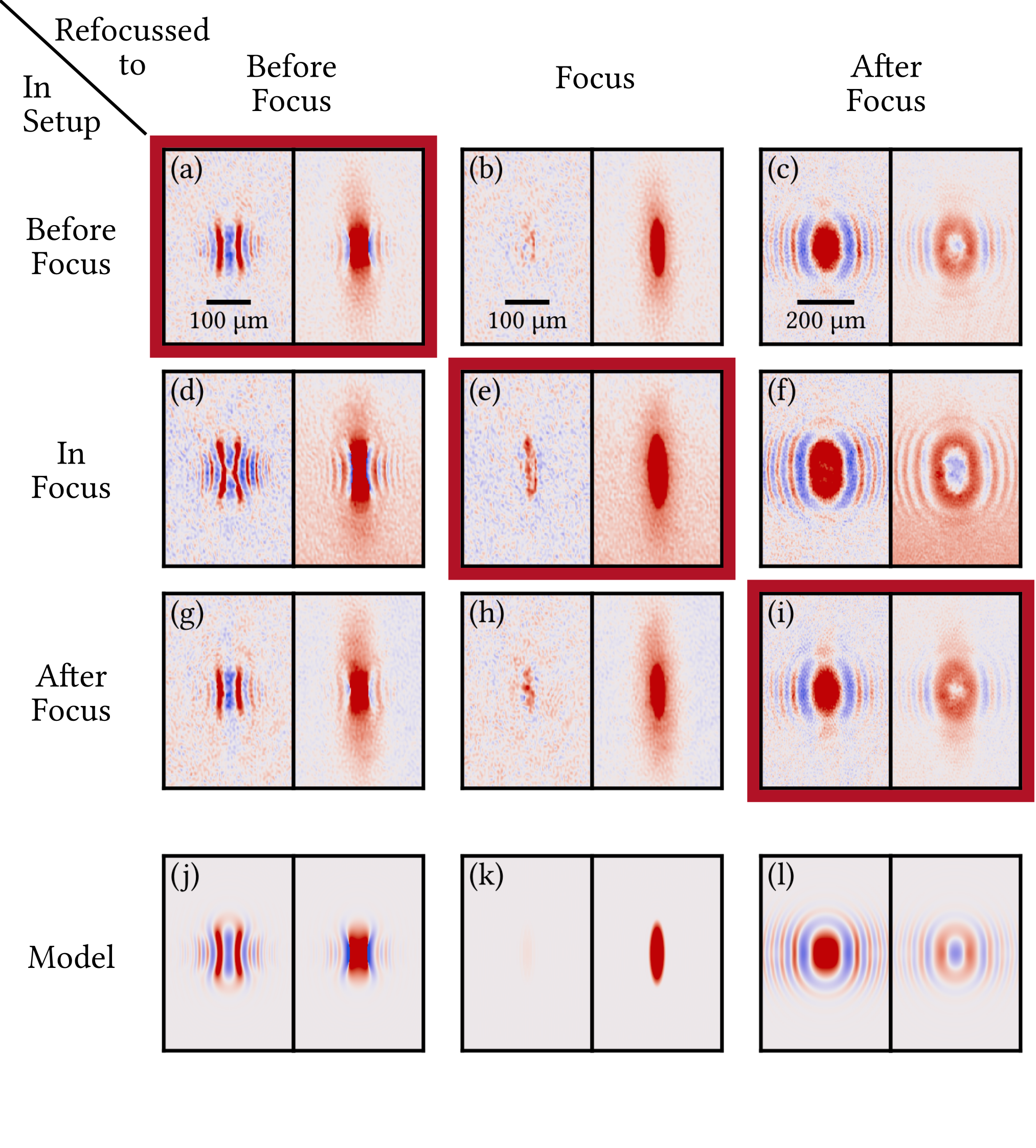}
\caption{Demonstration of numerical refocusing. For every image the left half shows the optical density, the right half shows the accumulated phase. Both are clipped at $\pm 1\,\mathrm{rad}$. Signal strength varies due to shot-to-shot variations in particle number. Rows represent different locations of the image plane in the setup: \emph{(a-c)} before the focus, \emph{(d-f)} in focus, \emph{(g-i)} after the focus, and \emph{(j-l)} comparison to the theoretical model. The columns represent a numerical refocusing to a certain image plane, chosen such that the diagonal, which is indicated by red borders, contains the data as recorded by the camera.}
\label{fig:refocusing}
\end{figure}

The information from Fig. \ref{fig:focus} can be used to accurately position the image plane in the same plane as the atom cloud. However, as the full field is measured, it is also possible to numerically propagate the image plane to the plane that contains the atoms, using Eq. \ref{eq:refocus} in 2D. To demonstrate this process of numerical refocusing, atom clouds are recorded with intentional defocusing by moving one of the imaging lenses in the setup. In Fig.~\ref{fig:refocusing} the result of this measurement is shown. The red boxes around frames (a), (e) and (i) indicate the actual images recorded in the experiment. Each row corresponds to a single measurement run and columns indicate different image planes, using data which is obtained by numerical propagation. For comparison, the BPM and a time-splitting spectral method has been used to calculate the expected intensity and phase for the parameters in Fig.~\ref{fig:refocusing}e at each image plane with a window function to attain the same numerical aperture as in the experiment. When the atom cloud is not in focus, a clear diffraction pattern is observed. Numerical refocusing of out-of-focus images reproduces both the thermal cloud and the Bose-Einstein condensate very accurately. Propagating to the image planes for the other two experiments yields similar diffraction patterns. This demonstrates that in our method the choice of image plane is irrelevant, since both phase and amplitude of the probe beam are known. In addition, in the case that more than one atom cloud is present, both can be imaged in a single shot, and then individually brought into focus numerically. Moreover, this method allows for correction of coma and spherical aberrations of the imaging system during the post-processing step in a manner similar to the treatment of refocusing here \cite{Faulkner03}.

In conclusion, we present a holographic method for imaging trapped quantum gases which provides significant advantages over established methods due to the inherent heterodyne gain. Using off-axis holography we retrieve the phase delay in the sample directly, as opposed to phase contrast imaging or shadowgraph imaging which yield a signal with a non-linear dependence on the phase and require subsequent unwrapping susceptible to noise. Due to the heterodyne gain, density profiles suitable for quantative analysis are obtained at probe doses two orders of magnitude smaller compared to phase-constrast imaging. This makes it possible to record hundreds of images of the same atom cloud, which enables the study of long term dynamics on a single sample. Using the phase and amplitude, the image plane is numerically scanned to determine the imaging plane of the atoms. Moreover, we demonstrate numerical refocusing in the case of defocusing, which also provides the possibility to correct coma and spherical aberrations of the imaging system. Converting an existing imaging system for quantum gases to use off-axis holography is done by adding a single reference beam to illuminate the camera, and therefore we envision the method will be adopted in every cold atoms laboratory.

\section*{Funding}
The work of A.P.M. is supported by the Netherlands Organisation for Scientific Research (NWO) through Vici grant 68047618.

\section*{Acknowledgements}
The authors thank T.\,S.~Loth and J.~Bosch for their contributions to the early stages of this research, and P.~Jurrius, C.R.~de~Kok, D.~Killian and F.~Ditewig for technical support.

\section*{Disclosures}
The authors declare no conflicts of interest.

\bibliography{atom_offaxis}


\bibliographyfullrefs{atom_offaxis}

\end{document}